\documentclass[lengthcheck,unsortedaddress,superscriptaddress,showpacs,letterpaper,nofootinbib]{revtex4-1}

\usepackage{comment}
\usepackage{color}
\usepackage{graphicx}  
\usepackage{float}
\usepackage{amsmath}
\usepackage{subfig}
\usepackage{acronym}
\usepackage{appendix}
\usepackage{placeins}
\usepackage[bookmarksnumbered, bookmarksopen, breaklinks, colorlinks]{hyperref}

\newcommand{\AEI}{Albert-Einstein-Institut, Max-Planck-Institut f\"ur
Gravitationsphysik, D-30167 Hannover, Germany}
\newcommand{\Leibniz}{Leibniz Universit\"at Hannover, D-30167 Hannover, Germany}

\begin{document}

\title{The Balancing Act Of Template Bank Construction:\\
\normalsize\textbf Inspiral Waveform Template Banks For Gravitational-Wave Detectors\\
And Optimizations At Fixed Computational Cost}

\author{Drew~Keppel}  
\email{drew.keppel@ligo.org}
\affiliation{\AEI}
\affiliation{\Leibniz}

\begin{abstract}

Gravitational-wave searches for signals from inspiralling compact binaries have
relied on matched filtering banks of waveforms (called \emph{template banks})
to try to extract the signal waveforms from the detector data. These template
banks have been constructed using four main considerations, the region of
parameter space of interest, the sensitivity of the detector, the matched
filtering bandwidth, and the sensitivity one is willing to lose due to the
granularity of template placement, the latter of which is governed by the
minimal match.  In this work we describe how the choice of the lower frequency
cutoff, the lower end of the matched filter frequency band, can be optimized
for detection. We also show how the minimal match can be optimally chosen in
the case of limited computational resources. These techniques are applied to
searches for binary neutron star signals that have been previously performed
when analyzing Initial LIGO and Virgo data and will be performed analyzing
Advanced LIGO and Advanced Virgo data using the expected detector sensitivity.
By following the algorithms put forward here, the volume sensitivity of these
searches is predicted to improve without increasing the computational cost of
performing the search.

\end{abstract}

\maketitle

\acrodef{BNS}{binary neutron star}
\acrodef{FAP}{false alarm probability}
\acrodef{FIR}{finite impulse response}
\acrodef{GW}{gravitational-wave}
\acrodef{PN}{post-Newtonian}
\acrodef{PSD}{power spectral density}
\acrodef{SNR}{signal-to-noise ratio}
\acrodef{SVD}{singular value decomposition}
\acrodef{aLIGO}{Advanced LIGO}
\acrodef{aVirgo}[AdV]{Advanced Virgo}

\section{Introduction}

For the past decade, large scale interferometric \ac{GW} detectors have
operated, allowing searches for signals from inspiralling compact binaries to
be performed~\cite{Tagoshi:2000bz, Takahashi:2004bj, CBC:S1, CBC:S2BNS,
CBC:S2MACHO, CBC:S2BBH, CBC:LIGOTAMA, CBC:S3S4, CBC:S3Spin, CBC:S51yr,
CBC:S5:12_18, CBC:S5LV, CBCHM:S5, CBC:S6, LIGO:CBC:S5:GRB, GRB070201,
GRB051103, LIGO:S6:GRB}. These searches have thus far detected no \ac{GW}
signals, however once the detectors are upgraded to their advanced
configurations, multiple events are expected to be detected each
year~\cite{rates2010}.

Searches for inspiral signals in detector data depend on matched filtering the
data with template waveforms to produce \ac{SNR} time-series, the maxima of
which are used to produce \ac{GW} ``triggers". Important criteria in
constructing banks of template waveforms (i.e., template banks) for these
searches are the region of parameter space to be searched, the sensitivity of
the detector, the lower and upper frequency cutoffs associated with matched
filtering the data, and the maximum fractional loss of \ac{SNR} (the complement
of which is more commonly know as the minimal match) that one is willing to
tolerate due to granularity of the template placement. Of these criteria, one
is free to tune the lower frequency cutoff and the minimal match due to
sensitivity and computational cost considerations.

In~\cite{Babak2006}, the authors discuss the issue of balancing computational
cost versus \ac{SNR} gain while decreasing the lower frequency cutoff. However,
they do not venture so far as to derive the optimal choices. Instead, they
choose to set the lower frequency cutoff at a level such that one would lose
less than 1\% of the \ac{SNR} by the cutoff being different from 0. In
addition, they choose the minimal match of the template bank to be $MM = 95\%$;
large enough that the metric estimate of the fractional \ac{SNR} loss is still
valid but small enough for computational cost considerations. Recent searches
for \ac{GW} from inspiralling compact binaries have chosen a larger value for
the minimal match, $MM = 97\%$, so that less than 10\% of the signals at the
worst mismatch locations of the template bank would be lost.

In this paper, we further investigate the effects of different lower frequency
cutoff and minimal match choices. In Sec.~\ref{sec:flow} we look at how
decreasing the lower frequency cutoff both increases the amount of raw \ac{SNR}
one is able to extract from a signal and increases the trials factor by
increasing the number of templates required to search for the waveforms.
Sec.~\ref{sec:cost} goes on to describe how to choose the optimal combination
of lower frequency cutoff and minimal match for a fixed computational cost.
Examples of both these choices are given in Sec.~\ref{sec:ex} where the methods
are applied to previous and future searches of \ac{GW} detector data.

\section{Preliminaries}
\label{sec:prelim}

In searching for signals from inspiralling compact objects in \ac{GW} data, a
commonly used event identification algorithm relies on matched filtering, where
the data is ``whitened" and filtered with the template waveform being searched
for.  Specifically, the square \ac{SNR} is given by
\begin{equation}\label{eq:snr}
\rho^2 = \frac{(s|h_c)^2 + (s|h_s)^2}{\sigma^2},
\end{equation}
where $s$ is the data from a detector that may contain a \ac{GW} signal of
unknown strength, $h_c$ and $h_s$ are the target waveforms associated with the
same source and differ in phase by $\pi/4$, $\sigma^2 := (h_c|h_c)$ is the
sensitivity of our detector to a waveform at a reference distance, typically
chosen to be 1 Mpc, and the inner product $(x|y)$ is defined as
\begin{equation}\label{eq:innerprod}
(x|y) := 4\Re\int_{f_{\rm low}}^{f_{\rm high}} \frac{\tilde{x}
\tilde{y}^*}{S_n(f)}df.
\end{equation}
Here $\tilde{x}$ is the Fourier transform of $x$, $()^*$ denotes the complex
conjugate operator, and $S_n(f)$ is the one-sided \ac{PSD} of the detector's
noise.

As can be seen from~\eqref{eq:innerprod}, the \ac{SNR} recovered when there is
a signal present in the data will depend on the limits of the integration. The
upper frequency cutoff $f_{\rm high}$ is set by the lower of either the Nyquist
frequency of the data or the maximum frequency of the template waveform. In
contrast, the lower frequency cutoff $f_{\rm low}$ is a parameter that can be
tuned in optimizing the search algorithm.

To search a region of parameter space, many template waveforms from points
spread throughout the region need to be matched filtered. The locations of
these points are chosen by constructing a metric on the parameter space
$g_{ij}$~\cite{Owen:1995tm, Owen:1998dk, Brown2012, Keppel2012}. This metric
describes the distance between points based on the fractional loss of \ac{SNR}
associated with matched filtering a signal waveform from one point in parameter
space with a template waveform from another point.  To second order in the
parameter differences $\Delta\lambda^i$, the fractional loss of \ac{SNR}, or
mismatch $m$, is given by
\begin{equation}
m = \frac{1}{2} g_{ij} \Delta\lambda^i \Delta\lambda^j,
\end{equation}
where the metric is given by projecting out dimensions of the parameter space
from normalized Fisher matrix,
\begin{equation}\label{eq:metric}
g_{\mu \nu} := \frac{(\partial_\mu h|\partial_\nu h)}{(h|h)},
\end{equation}
that are associated with extrinsic parameters, which can be maximized either
analytically or efficiently. Here $\partial_\mu$ is the partial derivative with
respect to parameter $\lambda^\mu$.  The density of templates is then governed
by the maximum amount of mismatch one is willing to tolerate, or the complement
of this, referred to as the \emph{minimal match} $MM = 1-m$.

\section{Signal Power versus Trials Factor:
Optimizing The Lower Frequency Cutoff For Maximum Sensitivity}
\label{sec:flow}

The goal of designing a search is to maximize the volume at which we are
sensitive to signals for a fixed \ac{FAP}. The first parameter we tune with
this in mind is the lower frequency cutoff. We start with the distance out to
which we can see an inspiral signal with a fixed \ac{SNR} $\rho$,
\begin{equation}\label{eq:dist}
D = \frac{\sigma}{\rho}.
\end{equation}
Changing the lower frequency cutoff changes the power of the signal that we
could possibly recover. If one were to recover a signal with the same \ac{SNR},
the distance to which one could see a signal would vary when the lower
frequency cutoff was changed from $f_{\rm ref}$ to $f_{\rm low}$,
\begin{equation}\label{eq:sigmadist}
\frac{D(f_{\rm low})}{D(f_{\rm ref})} = \frac{\sigma(f_{\rm
low})}{\sigma(f_{\rm ref})}.
\end{equation}

Let us now look at how the observable distance of a signal is affected when the
signal is recovered with a mismatched template. The observed \ac{SNR} $\rho$
will be reduced from the \ac{SNR} obtained by a template that matches the
signal $\rho_{\rm ref}$ by
\begin{equation}\label{eq:mismatchsnr}
\rho = \rho_{\rm ref} (1 - m),
\end{equation}
where $m$ is the mismatch between the template that recovers the signal and the
actual signal. Eq.~\ref{eq:dist} implies that the distance to which such a
signal will be observable is reduced by the same factor
\begin{equation}\label{eq:misdist}
\frac{D(f_{\rm ref},m)}{D(f_{\rm ref},0)} = (1 - m).
\end{equation}

So far we have focused on the obserable distance of a signal at fixed \ac{SNR}.
However it is actually the obserable distance of a signal at fixed \ac{FAP}
that we are interested in. The \ac{FAP} associated with a single observation of
\ac{SNR} $\rho$ is given by
\begin{equation}\label{eq:fapsnr}
\mathrm{FAP} \propto \exp[-\rho^2].
\end{equation}
The recovered \ac{FAP} is subject to a trials factor related to the number of
independent trials $N$ we use in looking for a signal,
\begin{equation}\label{eq:faptrialsfac}
\mathrm{FAP}' = 1 - (1 - \mathrm{FAP})^{N} \approx N \mathrm{FAP}.
\end{equation}
We can translate a single observation of $\rho_{\rm observed}$ among $N$
independent trials to a reference \ac{SNR} $\rho_{\rm ref}$ among a different
number of trials $N_{\rm ref}$ at the same \ac{FAP} by combining
\eqref{eq:fapsnr} and \eqref{eq:faptrialsfac}.
\begin{equation}\label{eq:trialssnr}
\rho_{\rm observed}^2 = \rho_{\rm ref}^2 + \ln\frac{N}{N_{\rm ref}}.
\end{equation}

When searching a non-zero measure region of parameter space, additional trials
are accrued proportional to the volume of the parameter space. The volume of
parameter space is in turn given by the number of templates needed to cover the
parameter space $M_{\rm templates}$~\cite{Prix2007},
\begin{equation}\label{eq:numtemps}
N_{\rm trials} \propto \int \sqrt{\left| g \right|} d\lambda^d = \frac{M_{\rm
templates} m^{d/2}}{\theta} 
\end{equation}
where $\sqrt{\left| g \right|}$ is the square root of the determinant of the
metric on the space, $\theta$ is a geometrical quantity associated with how the
template bank tiles the parameter space, $m = 1 - MM$ is maximum mismatch
allowed in the template bank covering the parameter space, and $d$ is the
dimensionality of the parameter space being tiled (i.e., two for templates
associated with waveforms from non-spinning objects that are laid out in the
two dimensional mass space).

Since the metric \eqref{eq:metric} is defined in terms of the inner products
from \eqref{eq:innerprod}, the full metric itself is a function of the lower
frequency cutoff, which in turn implies that the metric density of the mass
subspace is also a function of $f_{\rm low}$,
\begin{equation}\label{eq:metricdensityoff}
\sqrt{|g|} = \sqrt{|g(f_{\rm low})|}.
\end{equation}

The total volume we can observe is proportional to the cube of the distance,
thus the ratio of the volume we can observe for a mismatched signal at a given
value of $f_{\rm low}$ to the volume we could observe a matched signal with a
reference lower frequency cutoff $f_{\rm ref}$ is found by combining
\eqref{eq:sigmadist}, \eqref{eq:misdist}, \eqref{eq:trialssnr},
\eqref{eq:numtemps}, and \eqref{eq:metricdensityoff},
\begin{multline}
\frac{V(f_{\rm low},m)}{V(f_{\rm ref},0)} =  \frac{\sigma^3(f_{\rm
low})}{\sigma^3(f_{\rm ref})} \\
\times \frac{(1-m)^3}{\left( 1 + \frac{1}{\rho^2(f_{\rm
ref})}\ln\left[\frac{\int \sqrt{\left| g(f_{\rm low}) \right|} d\lambda^d}{\int
\sqrt{\left| g(f_{\rm ref}) \right|} d\lambda^d}\right]\right)^{3/2}}.
\end{multline}
We call this the relative volume. For two-dimensional template banks, a
hexagonal covering of templates following the $A_2^*$ lattice will result in a
distribution of mismatches that is essentially flat between 0 and the maximum
mismatch~\cite{Prix2007}. Using this fact, the average relative volume is found
to be
\begin{multline}\label{eq:relativevol}
\frac{\left \langle V(f_{\rm low},m) \right\rangle}{\left \langle V(f_{\rm
ref},0) \right\rangle} = \frac{\sigma^3(f_{\rm low})}{\sigma^3(f_{\rm ref})} \\
\times \frac{\left\langle \left( 1 - m \right)^3 \right\rangle}{\left( 1 +
\frac{1}{\rho^2(f_{\rm ref})}\ln\left[\frac{\int \sqrt{\left| g(f_{\rm low})
\right|} d\lambda^d}{\int \sqrt{\left| g(f_{\rm ref}) \right|}
d\lambda^d}\right]\right)^{3/2}},
\end{multline}
where the average of the mismatch term in the numerator is given by
\begin{equation}
\left\langle \left( 1 - m \right)^3 \right\rangle = 1 - \frac{3}{2}m + m^2 -
\frac{1}{4}m^3.
\end{equation}
The average relative volume can be maximized with the proper choice of $f_{\rm
low}$ for a fixed value of the template bank maximum mismatch.

\section{Wider or Denser?: Maximizing Sensitivity At Fixed Computational Cost}
\label{sec:cost}

In the face of limited computational resources, we must consider not only how
to maximize the sensitivity of a search through the choice of the lower
frequency cutoff, but we must also ensure that our choices of the lower
frequency cutoff and the minimal match satisfy the constraint on the total
computational cost $C_{\rm total}$. This constraint can be viewed as a
combination of two effects: the computational cost of filtering the data with a
single template waveform $C_{\rm filter}$ multiplied by the computational cost
associated with $N_{\rm templates}$ such filters
\begin{align}
C_{\rm total}(f_{\rm low}) =& N_{\rm templates}(f_{\rm low}) C_{\rm
filter}(f_{\rm low}) \nonumber \\
=& C_{\rm filter}(f_{\rm low}) \theta m^{-d/2} \int \sqrt{\left| g(f_{\rm low})
\right|} d\lambda^d.
\end{align}
Using this constraint, we seek to maximize the constrained average relative 
volume
\begin{align}\label{eq:costrelavevol}
\frac{\left \langle V(f_{\rm low}) \right\rangle}{\left \langle V(f_{\rm ref})
\right\rangle} =& \frac{\sigma^3(f_{\rm low})}{\sigma^3(f_{\rm ref})}
\frac{\left\langle \left( 1 - m(f_{\rm low}) \right)^3
\right\rangle}{\left\langle \left( 1 - m(f_{\rm ref}) \right)^3 \right\rangle}
\nonumber\\ &\times\frac{1}{\left( 1 + \frac{1}{\rho^2(f_{\rm
ref})}\ln\left[\frac{\int \sqrt{\left| g(f_{\rm low}) \right|} d\lambda^d}{\int
\sqrt{\left| g(f_{\rm ref}) \right|} d\lambda^d}\right]\right)^{3/2}},
\end{align}
with the proper choice of $f_{\rm low}$ and $m(f_{\rm low})$.

Assuming one is able to computationally preform the search for a given
combination of lower frequency cutoff $f_{\rm ref}$ and maximum mismatch
$m(f_{\rm ref})$, the maximum mismatch at any other choice of lower frequency
cutoff $f_{\rm low}$ satisfying the constraint on the computational cost can be
solved for easily,
\begin{equation}\label{eq:matc}
m(f_{\rm low}) = m(f_{\rm ref}) \left( \frac{C_{\rm filter}(f_{\rm low}) \int
\sqrt{\left| g(f_{\rm low}) \right|} d\lambda^d}{C_{\rm filter}(f_{\rm ref})
\int \sqrt{\left| g(f_{\rm ref}) \right|} d\lambda^d} \right)^{2/d}
\end{equation}

The computational cost of filtering data with a single template will depend
intrinsically on the implementation of a search.  As a first example, it could
be independent of the choice of $f_{\rm low}$, as is the case in the
\texttt{FINDCHIRP} algorithm~\cite{findchirppaper} where data is processed with
fast Fourier transforms using fixed length chunks.

In a different algorithm where data is analyzed in the time domain using
\ac{FIR} filters, the computational cost would be set by the number of taps in
the \ac{FIR} filter.  This is proportional to the length of the waveform $T$,
given to Newtonian order by
\begin{equation}\label{eq:wavelength}
T(f_{\rm low}) = \frac{5}{256 \mathcal{M}^{5/3}} \left[(4\pi f_{\rm
low})^{-8/3} - (4\pi f_{\rm high})^{-8/3}\right],
\end{equation}
where $\mathcal{M} = (m_1 m_2)^{3/5} (m_1 + m_2)^{1/5}$ is the chirp mass of
the binary system.

Alternatively, if one were able to change the sampling rate associated with the
template filter continuously, one could reduce the computational cost by
filtering the data with a changing local sampling rate such that the frequency
of the signal at any time was always equal to the local Nyquist frequency of
the filter. In this approach, the computational cost would be proportional to
the number of cycles in the signal waveform,
\begin{equation}\label{eq:cyclescost}
N_{\rm cycles}(f_{\rm low}) = \frac{1}{64 \pi^{8/3} \mathcal{M}^{5/3}}\left(
f_{\rm low}^{-5/3} - f_{\rm high}^{-5/3} \right).
\end{equation}

Finally, as an application of this method to a  pipeline proposed to search for
\ac{BNS} signals with low latency in the \ac{aLIGO} sensitive band, we consider
the computational cost of the \texttt{LLOID} algorithm~\cite{Cannon2011Early}.
The \texttt{LLOID} algorithm partitions the waveforms into $S$ time-slices and
filters the waveform portions of slice $s$ at a power of two sampling rate
$f^s$ such that the Nyquist frequency of the slice is just greater than the
largest frequency of any of the portions of the waveform in that slice. In
addition, for each slice, the \texttt{LLOID} algorithm decomposes the $N_{\rm
templates}$ template waveform portions into $L_{\rm bases}^s$ basis vectors
using \ac{SVD}~\cite{Cannon2010}.  These basis vectors are used as \ac{FIR}
filters of $N^s_{\rm taps}$ taps for slice $s$. The computational cost of
filtering a bank of waveforms with this algorithm is dominated by the filtering
costs of the basis vectors and the reconstruction costs of turning the basis
filter outputs into outputs of template filters,
\begin{equation}\label{eq:lloidcost}
N_{\rm FLOPS} = 2\sum_{s=0}^{S-1} f^s L^s_{\rm bases} (N_{\rm taps}^s + N_{\rm
templates}).
\end{equation}

Let us look at how the different pieces of \texttt{LLOID}'s computational cost
will change with varying $f_{\rm low}$. For a particular slice, as $f_{\rm
low}$ is reduced, the template waveforms that go into the \ac{SVD} matrix will
more densely cover the region of parameter space, resulting in a larger number
of waveforms that will need to be reconstructed (i.e., $N_{\rm templates}$ will
increase). However, the number of bases $N^s_{\rm bases}$ needed reconstruct
the template waveforms to a specific accuracy is invariant for the minimal
matches we are interested in~\cite{Cannon2011manifold}. Finally, the number of
slices kept will depend on $f_{\rm low}$ as each slice covers a different
frequency range of the waveforms. Thus, the total computational cost of the
\texttt{LLOID} algorithm can be written as
\begin{equation}
N_{\rm FLOPS} = A(f_{\rm low}) N_{\rm templates} + B(f_{\rm low}),
\end{equation}
where $A(f_{\rm low})$ and $B^s(f_{\rm low})$ are defined appropriately with
respect to \eqref{eq:lloidcost}. For this algorithm, \eqref{eq:matc} takes a
different form,
\begin{equation}
m(f_{\rm low}) = m(f_{\rm ref}) \left( \frac{A(f_{\rm low}) N_{\rm ref}
\frac{\int \sqrt{\left| g(f_{\rm low}) \right|} d\lambda^d}{\int \sqrt{\left|
g(f_{\rm ref}) \right|} d\lambda^d}}{A(f_{\rm ref})N_{\rm ref} + B(f_{\rm ref})
- B(f_{\rm low})} \right)^{2/d},
\end{equation}
where $N_{\rm ref} := N_{\rm templates}(f_{\rm low})$.

\FloatBarrier
\section{Examples}
\label{sec:ex}

\begin{figure}
\centering
\subfloat[H1 PSDs]{\includegraphics[]{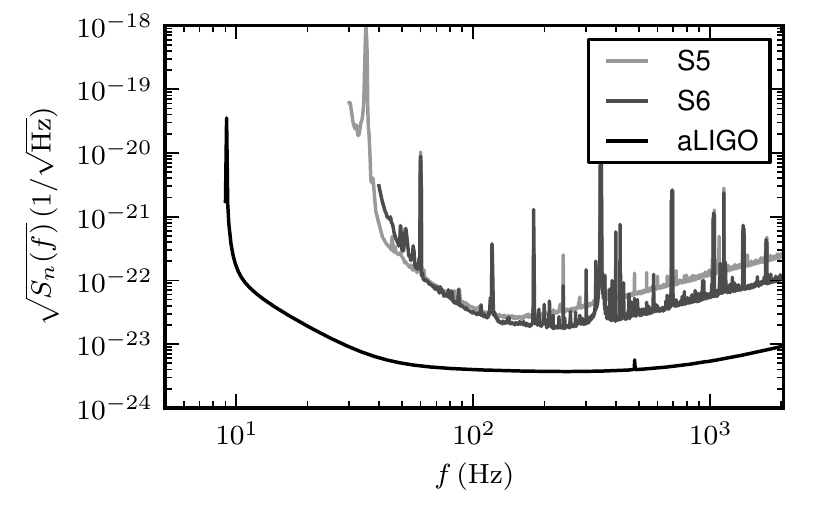}\label{fig:H1psds}}
\qquad
\subfloat[V1 PSDs]{\includegraphics[]{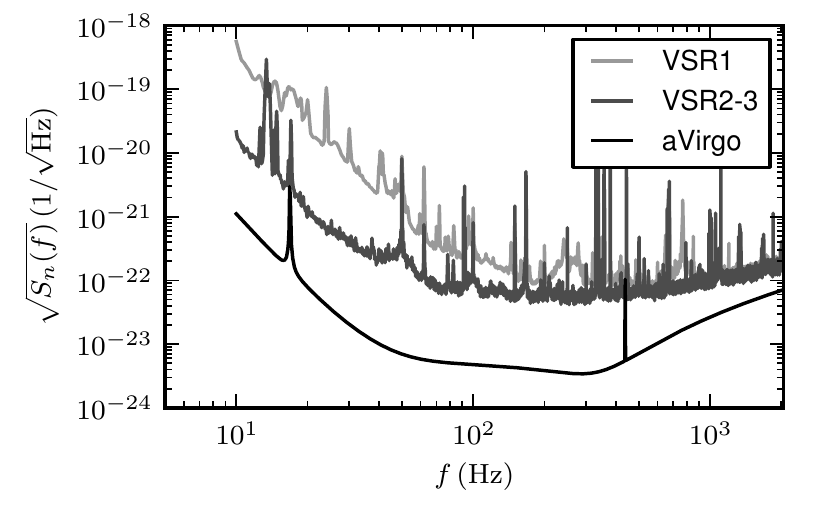}\label{fig:V1psds}}
\qquad
\subfloat[Harmonic Sum PSDs]{\includegraphics[]{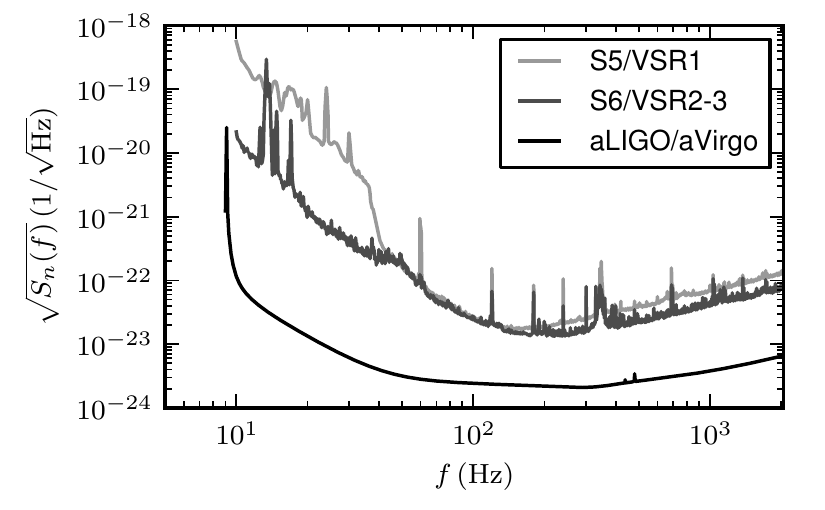}\label{fig:harmonicpsds}}
\caption{\protect\subref{fig:H1psds} shows different \acp{PSD} associated with
different eras of the H1 LIGO detector. \protect\subref{fig:V1psds} shows different
\acp{PSD} associated with different eras of the Virgo detector.
\protect\subref{fig:harmonicpsds} shows different \acp{PSD} associated with detector
networks from different eras. The H1H2L1V1 network \ac{PSD} associated with the
S5/VSR1 era is given by the harmonic sum of individual detectors' \acp{PSD}.
For the S6/VSR2-3 and \ac{aLIGO}/\ac{aVirgo} eras, an H1L1V1 network is used.}
\label{fig:psds}
\end{figure}

In this section we apply the methods from Secs.~\ref{sec:flow} and
\ref{sec:cost} to the (expected) sensitivities of several past and future
detectors.  In particular, we investigate the LIGO and Virgo \acp{PSD} from
S5/VSR1~\cite{Abadie:2010cg}, S6/VSR2-3~\cite{LIGO:2012aa, Accadia:2010aa}, and
the expected advanced detector \acp{PSD} for \ac{aLIGO}~\cite{advLIGO} and
\ac{aVirgo}~\cite{advVirgoPSD}. We also consider joint detector analyses where
the individual detectors \acp{PSD} are combined by taking the harmonic sum,
which yields the same combined \ac{SNR} as either the coherent network \ac{SNR}
or the sum-of-squares \ac{SNR} associated with a coincident
search~\cite{Harry2011, Keppel2012coh2}. These \acp{PSD} can be seen in
Fig.~\ref{fig:psds}. The parameter space we focus on for these comparisons is
that associated with searches for \ac{BNS} signals from non-spinning objects.
Using the stationary phase approximation, we expand the template waveforms to
Newtonian order in the amplitude and 3.5~\ac{PN} order in the phase. The metric
for these waveforms is given in~\cite{Keppel2012}. With this focus, we
approximate the ratio of the integrated metric density by a point estimate such
that the mass of each object is 1.4 $M_\odot$,
\begin{equation}
\frac{\int \sqrt{\left| g(f_{\rm low}) \right|} d\lambda^d}{\int \sqrt{\left|
g(f_{\rm ref}) \right|} d\lambda^d} \approx \sqrt{\frac{\left| g_{\rm
BNS}(f_{\rm low}) \right|}{\left| g_{\rm BNS}(f_{\rm ref}) \right|}}.
\end{equation}
It should also be noted that, in this approximation, we assume that the effects
from the bulk of parameter space dominate over effects from the boundaries.
For parameter spaces where the effects of the boundaries are non-negligible,
more care will be needed in computing the ratio of the integrated metric
densities and how they relate to the trials factor and computational cost.

\FloatBarrier
\subsection{Choice of $f_{\rm low}$}
\label{subsec:exflow}

\begin{table}[t]
\centering
\begin{tabular}{c c c c c}
\hline\hline
Era & Detector & $f^{\textrm{standard}}_{\rm low}$ & $f^{\textrm{optimal}}_{\rm low}$ & Volume Gain \\ [0.5ex]
\hline
S5 & H1 & 40Hz & 37.3Hz & $6.4 \times 10^{-5}$ \\
VSR1 & V1 & 60Hz & 38.1Hz & $1.9 \times 10^{-2}$ \\
S5/VSR1 & H1H2L1V1 & 40Hz & 37.8Hz &  $4.9 \times 10^{-5}$ \\
S6 & H1 & 40Hz & 43.7Hz & $2.5 \times 10^{-5}$ \\
VSR2-3 & V1 & 50Hz & 16.8Hz & $1.5 \times 10^{-1}$ \\
S6/VSR2-3 & H1L1V1 & 40Hz & 34.0Hz & $7.0 \times 10^{-4}$ \\
\ac{aLIGO} & H1 & 10Hz & 9.6Hz & $1.3 \times 10^{-5}$ \\
\ac{aVirgo} & V1 & 10Hz & 17.6Hz & $1.7 \times 10^{-3}$ \\
\ac{aLIGO}/\ac{aVirgo} & H1L1V1 & 10Hz & 10.1Hz &  $8.6 \times 10^{-7}$ \\ [1ex]
\hline
\end{tabular}
\caption{We show the increase in the average relative volume
\eqref{eq:relativevol} that can be achieved by switching from the standard
lower frequency cutoff to the optimal lower frequency cutoff. The minimal match
in either case is set to be 3\%. The volume increase compared to the standard
choice is very small, except for the V1 VSR1 PSD, where a higher than normal
lower frequency cutoff was employed.}
\label{tbl:f_low_std}
\end{table}

First we optimize the choice of the lower frequency cutoff of an inspiral
search for different detectors without regard to the computational cost.
Table~\ref{tbl:f_low_std} summarizes the results for all of the detector
combinations mentioned, compared to the ``standard" choice of the lower
frequency cutoff. For the most part, this is a very small effect, as can be
anticipated through the logarithmic dependence of the effect of the trials
factor in~\eqref{eq:relativevol}. The largest differences between the standard
choice and the optimal choice occur for the Virgo detector during VSR1, which
increases the sensitivity of the search by $15\%$. This seems to be attributed
to a rapid decrease in the recoverable \ac{SNR} that is seen between about 55Hz
and 60Hz.

\begin{table}[h]
\centering
\begin{tabular}{c c c c c}
\hline\hline
Era & Detector & $f^{\textrm{minimum}}_{\rm low}$ & $f^{\textrm{optimal}}_{\rm low}$ & Volume Gain \\ [0.5ex]
\hline
S5 & H1 & 30Hz & 37.3Hz & $2.0 \times 10^{-6}$ \\
VSR1 & V1 & 10Hz & 38.1Hz & $3.5 \times 10^{-5}$ \\
S5/VSR1 & H1H2L1V1 & 10Hz & 37.8Hz &  $1.7 \times 10^{-4}$ \\
S6 & H1 & 40Hz & 43.7Hz & $2.5 \times 10^{-5}$ \\
VSR2-3 & V1 & 10Hz & 16.8Hz & $1.1 \times 10^{-3}$ \\
S6/VSR2-3 & H1L1V1 & 10Hz & 34.2Hz & $5.5 \times 10^{-3}$ \\
\ac{aLIGO} & H1 & 9Hz & 9.6Hz & $2.5 \times 10^{-6}$ \\
\ac{aVirgo} & V1 & 10Hz & 17.6Hz & $1.7 \times 10^{-3}$ \\
\ac{aLIGO}/\ac{aVirgo} & H1L1V1 & 9Hz & 10.1Hz &  $2.2 \times 10^{-5}$ \\ [1ex]
\hline
\end{tabular}
\caption{Similar to Table~\ref{tbl:f_low_std}, we show the increase in the
average relative volume \eqref{eq:relativevol} that can be achieved by
switching to the optimal lower frequency cutoff. However, here the reference
lower frequency cutoff is set to the minimum frequency at which a detector's
\ac{PSD} is reported.}
\label{tbl:f_low_min}
\end{table}

\begin{figure}
\centering
\subfloat[H1 S5 PSD]{\includegraphics[]{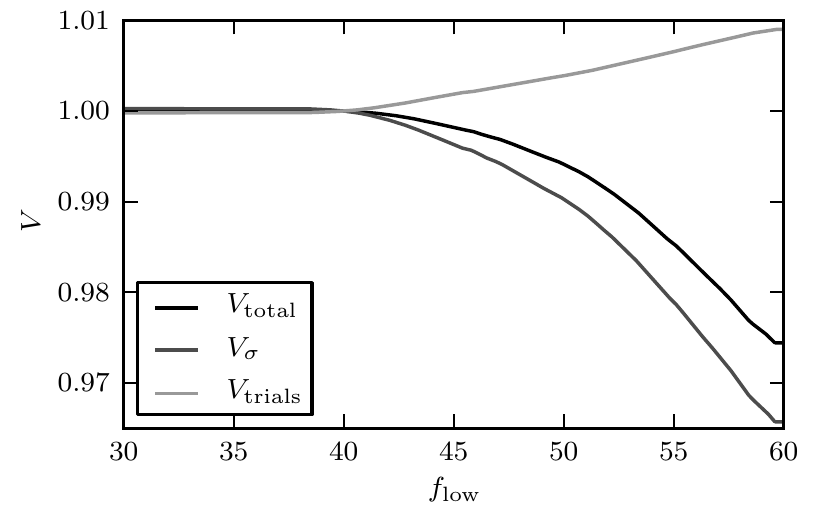}\label{fig:H1S5mm}}
\qquad
\subfloat[V1 VSR1 PSD]{\includegraphics[]{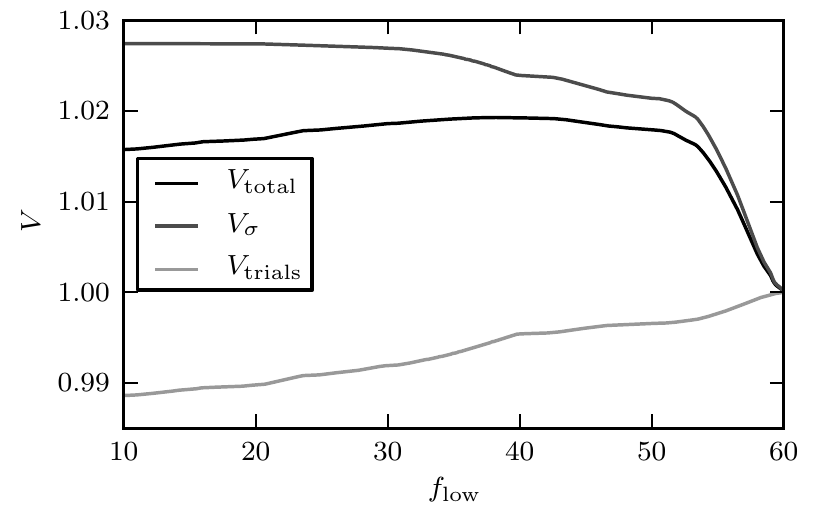}\label{fig:V1VSR1mm}}
\qquad
\subfloat[H1H2L1V1 S5/VSR1 Harmonic Sum PSD]{\includegraphics[]{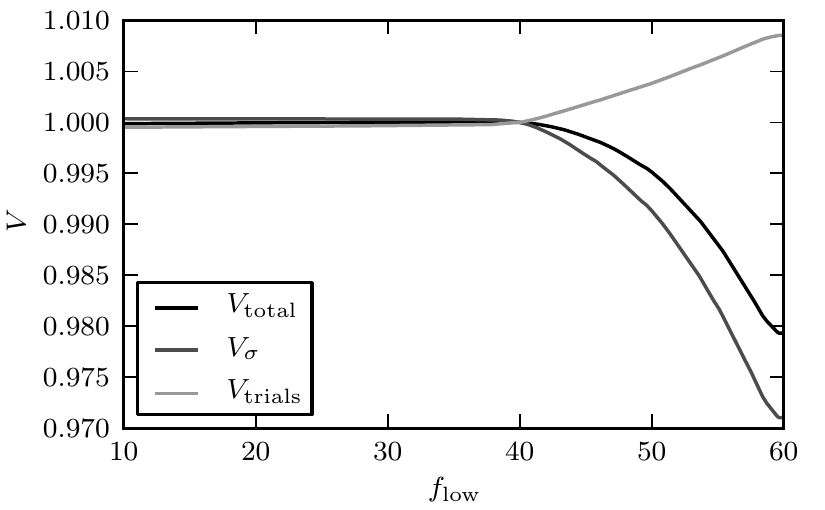}\label{fig:H1H2L1V1S5VSR1mm}}
\caption{\protect\subref{fig:H1S5mm} shows the average relative volume $V_{\rm total}$
as a function of lower frequency cutoff for the H1 LIGO detector during the S5
era.  \protect\subref{fig:V1VSR1mm} and \protect\subref{fig:H1H2L1V1S5VSR1mm} show the same for
the Virgo detector and H1H2L1V1 detector network for the VSR1 and S5/VSR1 eras,
respectively. Each panel also contains traces for the contributions to the
average relative volume from the recoverable \ac{SNR} $V_{\sigma}$ and the
trials factor $V_{\rm trials}$.}
\label{fig:S5VSR1mm}
\end{figure}

\begin{figure}
\centering
\subfloat[H1 S6 PSD]{\includegraphics[]{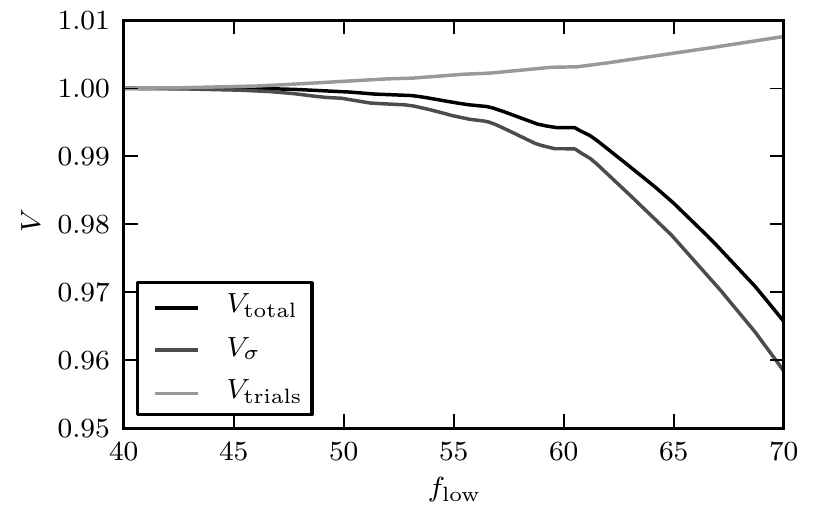}\label{fig:H1S6mm}}
\qquad
\subfloat[V1 VSR2-3 PSD]{\includegraphics[]{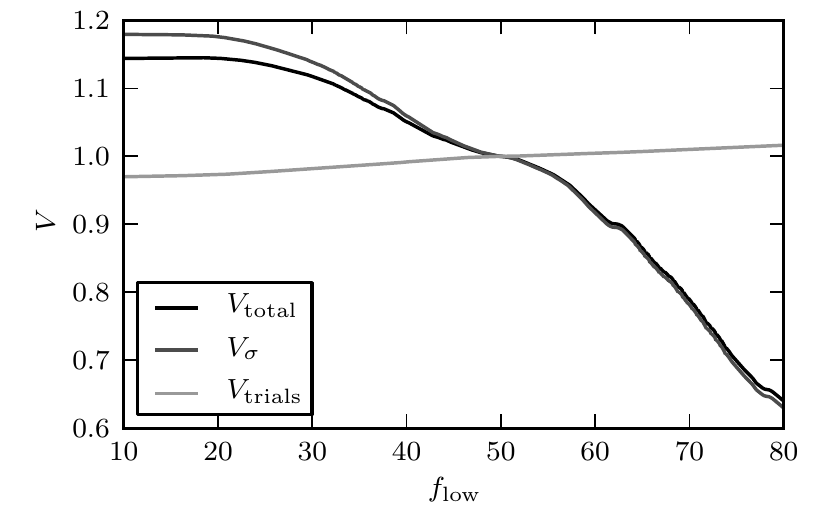}\label{fig:V1VSR23mm}}
\qquad
\subfloat[H1L1V1 S6/VSR2-3 Harmonic Sum PSD]{\includegraphics[]{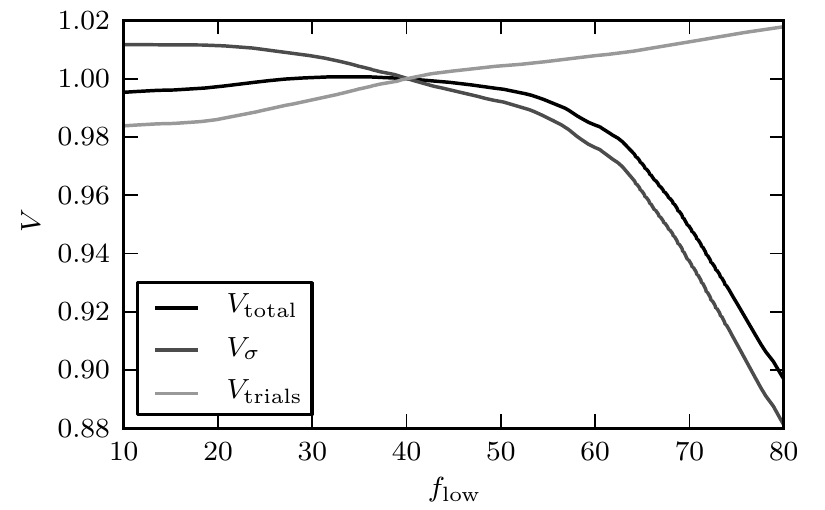}\label{fig:H1L1V1S6VSR23mm}}
\caption{\protect\subref{fig:H1S6mm} shows the average relative volume as a function of
lower frequency cutoff for the H1 LIGO detector during the S6 era.
\protect\subref{fig:V1VSR23mm} and \protect\subref{fig:H1L1V1S6VSR23mm} show the same for the
Virgo detector and H1L1V1 detector network for the VSR2-3 and S6/VSR2-3 eras,
respectively.  Each panel also contains traces for the contributions to the
average relative volume from the recoverable \ac{SNR} $V_{\sigma}$ and the
trials factor $V_{\rm trials}$.}
\label{fig:S6VSR23mm}
\end{figure}

\begin{figure}
\centering
\subfloat[H1 \ac{aLIGO} PSD]{\includegraphics[]{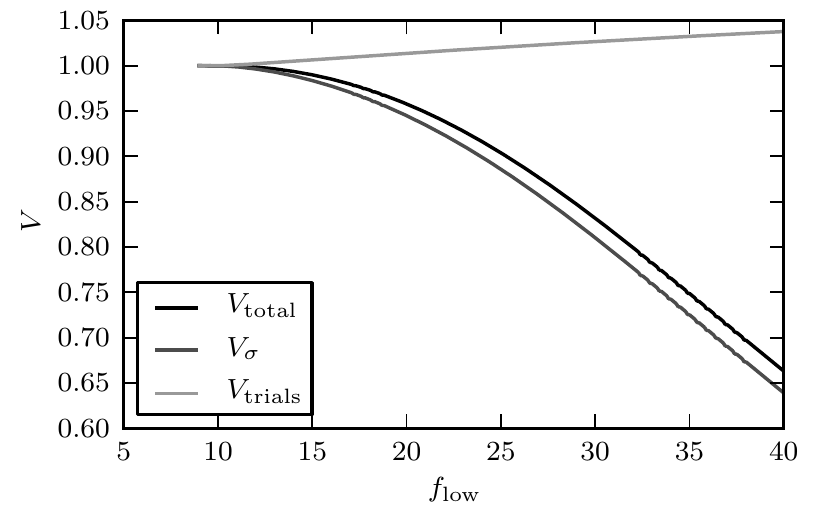}\label{fig:H1aLIGOmm}}
\qquad
\subfloat[V1 \ac{aVirgo} PSD]{\includegraphics[]{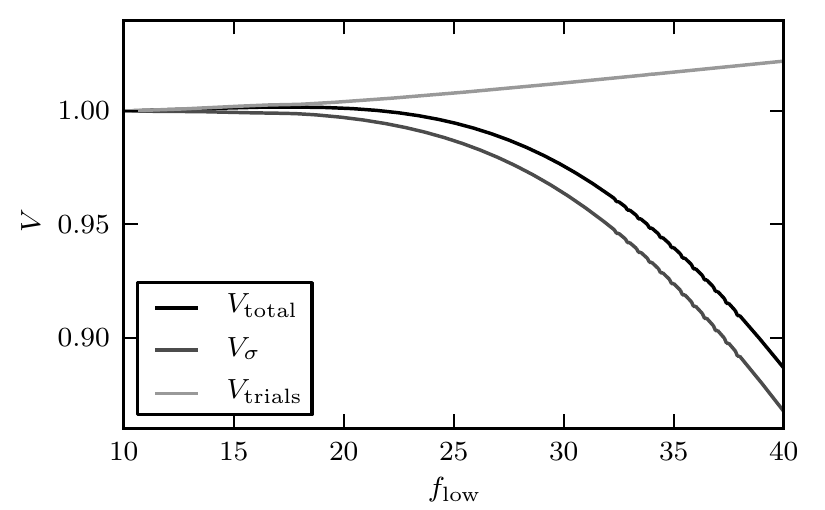}\label{fig:V1aVirgomm}}
\qquad
\subfloat[H1L1V1 \ac{aLIGO}/\ac{aVirgo} Harmonic Sum PSD]{\includegraphics[]{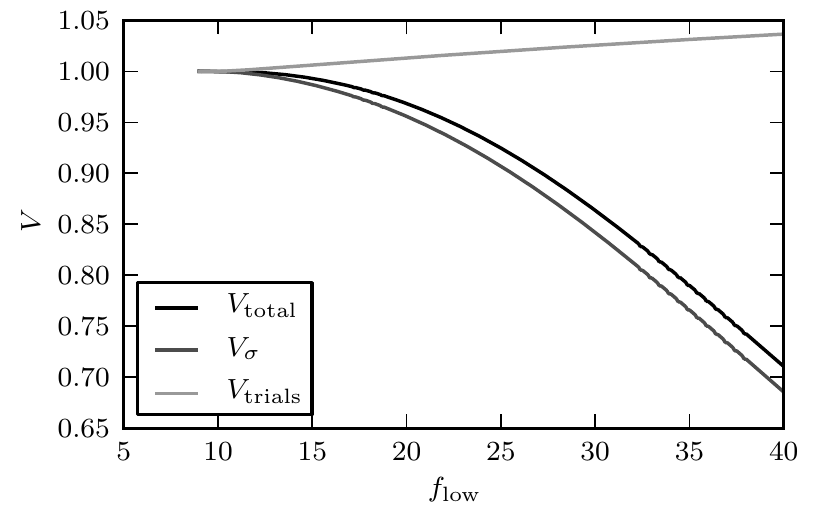}\label{fig:H1L1V2aLIGOaVirgomm}}
\caption{\protect\subref{fig:H1aLIGOmm} shows the average relative volume as a function
of lower frequency cutoff for the proposed H1 LIGO detector during the
\ac{aLIGO} era.  \protect\subref{fig:V1aVirgomm} and \protect\subref{fig:H1L1V2aLIGOaVirgomm}
show the same for the proposed Virgo detector and H1L1V1 detector network for
the \ac{aVirgo} and \ac{aLIGO}/\ac{aVirgo} eras, respectively.  Each panel also
contains traces for the contributions to the average relative volume from the
recoverable \ac{SNR} $V_{\sigma}$ and the trials factor $V_{\rm trials}$.}
\label{fig:aLIGOaVirgomm}
\end{figure}

Table~\ref{tbl:f_low_min} makes a similar comparison, although here the
standard lower frequency cutoff choice is replaced by the minimum reported
frequency associated with a particular \ac{PSD}. It is particularly interesting
to see the trials factor effect associated with the Virgo detector during VSR1.
In that case, the difference between the minimum choice of 10Hz and the optimal
choice of 38.1Hz is a few parts in $10^5$. What is interesting about this
comparison is the large difference in the lower frequency cutoff choices. As
Virgo detector's \ac{PSD} from VSR1 had a very shallow slope at the low
frequency end, it provides a good example of how the effect of the trials
factor can grow more quickly than the \ac{SNR} gain as the lower frequency
cutoff is lowered.  More detailed sensitivity comparisons can be found in
Figs.~\ref{fig:S5VSR1mm}-\ref{fig:aLIGOaVirgomm}, which separately show the
effect of varying lower frequency cutoff on the recovered \ac{SNR} and on the
trials factor effect as a function of the lower frequency cutoff. The example
described above associated with the Virgo VSR1 \ac{PSD} can be seen in
Fig.~\ref{fig:V1VSR1mm}.

\FloatBarrier
\subsection{Fixed Computational Cost}
\label{subsec:exfixedcost}

\begin{table*}
\centering
\begin{tabular}{c c c c c c}
\hline\hline
Detector & Era & Cost & $f^{\textrm{standard}}_{\rm low}$,
$m^{\textrm{standard}}_{\rm max}$ & $f^{\textrm{optimal}}_{\rm low}$,
$m^{\textrm{optimal}}_{\rm max}$ & Volume Gain \\ [0.5ex]
\hline
S5 & H1 & Fixed & 40Hz, 3\% & 49.1Hz, 2.4\% & $4.5 \times 10^{-3}$ \\
VSR1 & V1 & Fixed & 60Hz, 3\% & 50.1Hz, 2.8\% & $6.2 \times 10^{-3}$ \\
S5/VSR1 & H1H2L1V1 & Fixed & 40Hz, 3\% & 50.2Hz, 2.3\% &  $5.3 \times 10^{-3}$ \\
S6 & H1 & Fixed & 40Hz, 3\% & 55.7Hz, 2.6\% & $3.5 \times 10^{-3}$ \\
VSR2-3 & V1 & Fixed & 50Hz, 3\% & 37.8, 6.2\% & $1.8 \times 10^{-2}$ \\
S6/VSR2-3 & H1L1V1 & Fixed & 40Hz, 3\% & 51.2Hz, 2.2\% & $7.9 \times 10^{-3}$ \\
\ac{aLIGO} & H1 & Cycles & 10Hz, 3\% & 14.4Hz, 1.1\% & $2.1 \times 10^{-2}$ \\
\ac{aVirgo} & V1 & Cycles & 10Hz, 3\% & 22.0Hz, 0.56\% & $3.7 \times 10^{-2}$ \\
\ac{aLIGO}/\ac{aVirgo} & H1L1V1 & Cycles & 10Hz, 3\% & 15.0Hz, 1.0\% &  $2.3 \times 10^{-2}$ \\
\ac{aLIGO}/\ac{aVirgo} & H1L1V1 & LLOID & 9.7Hz, 3\% & 14.2Hz, 0.84\% &  $2.8 \times 10^{-2}$ \\ [1ex]
\hline
\end{tabular}
\caption{We show the gain in the constrained average relative volume
\eqref{eq:costrelavevol} that can be obtained by changing from the standard
choice of lower frequency cutoff and maximum mismatch to the optimal choice. The
computational cost for each of these calculations is set using the algorithm
listed under ``Cost". Most of these searches are optimized by increasing the
lower frequency cutoff and decreasing the maximum mismatch (i.e., increasing the
density) of the template bank.}
\label{tbl:costf_low}
\end{table*}

\begin{table*}
\centering
\begin{tabular}{c c c c c c}
\hline\hline
Era & Detector & Cost & $f^{\textrm{previous}}_{\rm low}$,
$m^{\textrm{previous}}_{\rm max}$ & $f^{\textrm{optimal}}_{\rm low}$,
$m^{\textrm{optimal}}_{\rm max}$ & Volume Gain \\ [0.5ex]
\hline
S5 & H1 & Fixed & 57.8Hz, 5\% & 58.4Hz, 4.9\% & $5.3 \times 10^{-5}$ \\
VSR1 & V1 & Fixed & 60.0Hz, 5\% & 50.1Hz, 5.4\% & $6.8 \times 10^{-4}$ \\
S5/VSR1 & H1H2L1V1 & Fixed & 60.0Hz, 5\% & 59.7Hz, 5.0\% &  $3.2 \times 10^{-7}$ \\
S6 & H1 & Fixed & 67.1Hz, 5\% & 63.2Hz, 5.7\% & $2.0 \times 10^{-3}$ \\
VSR2-3 & V1 & Fixed & 44.7Hz, 5\% & 43.4Hz, 5.2\% & $3.0 \times 10^{-4}$ \\
S6/VSR2-3 & H1L1V1 & Fixed & 62.8Hz, 5\% & 60.3Hz, 5.3\% & $6.7 \times 10^{-4}$ \\
\ac{aLIGO} & H1 & Cycles & 17.0Hz, 5\% & 19.5Hz, 3.1\% & $9.7 \times 10^{-3}$ \\
\ac{aVirgo} & V1 & Cycles & 28.8Hz, 5\% & 31.3Hz, 3.7\% & $5.1 \times 10^{-3}$ \\
\ac{aLIGO}/\ac{aVirgo} & H1L1V1 & Cycles & 18.0Hz, 5\% & 20.8Hz, 3.1\% & $1.1 \times 10^{-2}$ \\
\hline
\end{tabular}
\caption{We show the gain in the constrained average relative volume
\eqref{eq:costrelavevol} that can be obtained by changing from the choice of
lower frequency cutoff and maximum mismatch proposed in \cite{Babak2006} to the
optimal choice. Again, the computational cost for each of these calculations is
set using the algorithm listed under ``Cost". The choices of \cite{Babak2006}
are close to optimal, although the advanced detector network search can be
improved by of order one percent when switching to the optimal choices.}
\label{tbl:costoldf_low}
\end{table*}

We now consider the task of choosing optimal values for both the lower
frequency cutoff and the minimal match of the template bank subject to the
constraint of fixed computational cost. Table~\ref{tbl:costf_low} shows a
comparison between the standard values and the optimal values chosen using the
algorithm proposed in this paper. In addition to the detector/era
associated with a particular \ac{PSD} and the standard and optimal choices for
the minimal match and lower frequency cutoff, this table also lists the
computational cost algorithm that is appropriate for a given search.

We see that including the constraint on the computational cost produces a
larger effect than optimizing the lower frequency cutoff alone without the
constraint. It is interesting to note that for the majority of the cases
investigated, the optimal choice involves reducing the computational cost
through raising the lower frequency cutoff and then reinvesting the
computational savings into increasing the density of the template bank.

\begin{figure}[h]
\centering
\subfloat[H1 S5 PSD]{\includegraphics[]{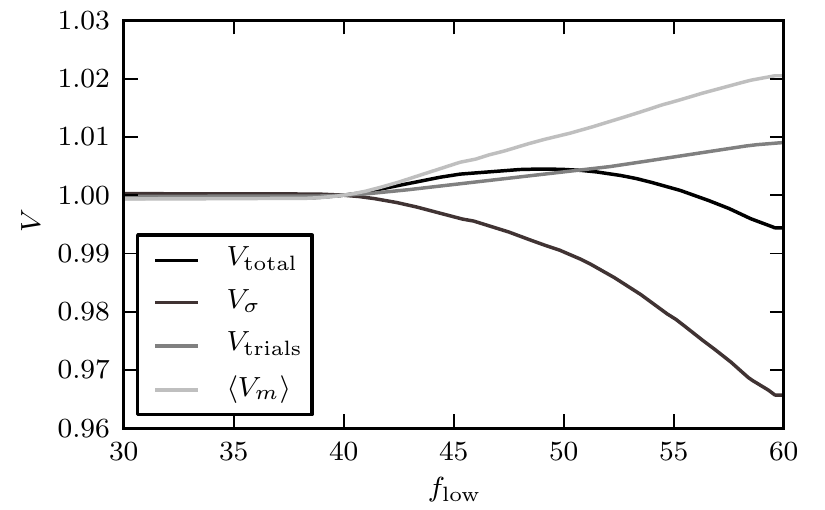}\label{fig:H1S5cost}}
\qquad
\subfloat[V1 VSR1 PSD]{\includegraphics[]{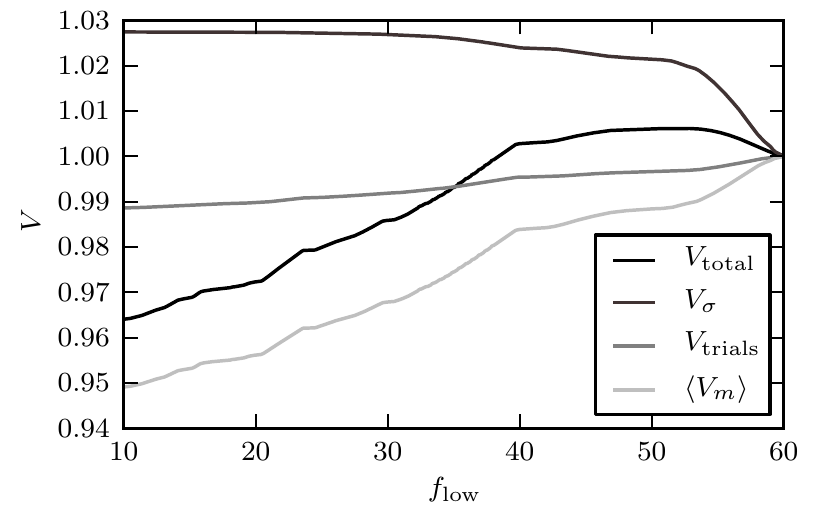}\label{fig:V1VSR1cost}}
\qquad
\subfloat[H1H2L1V1 S5/VSR1 Harmonic Sum PSD]{\includegraphics[]{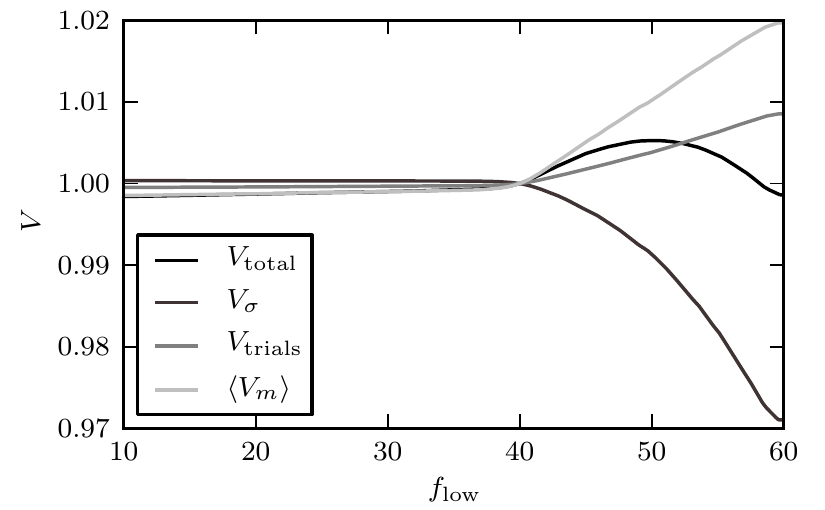}\label{fig:H1H2L1V1S5VSR1cost}}
\caption{\protect\subref{fig:H1S5cost} shows the constrained average relative volume as
a function of lower frequency cutoff for the H1 LIGO detector during the S5 era.
\protect\subref{fig:V1VSR1cost} and \protect\subref{fig:H1H2L1V1S5VSR1cost} show the same for
the Virgo detector and H1H2L1V1 detector network for the VSR1 and S5/VSR1 eras,
respectively. The computational cost of the searches associated with these
eras is given by the fixed cost algorithm. Each panel also contains traces for
the contributions to the constrained average relative volume from the
recoverable \ac{SNR}~$V_{\sigma}$, the trials factor~$V_{\rm trials}$, and the
average template bank mismatch~$\langle V_{m} \rangle$.}
\label{fig:S5VSR1cost}
\end{figure}

\begin{figure}[h]
\centering
\subfloat[H1 S6 PSD]{\includegraphics[]{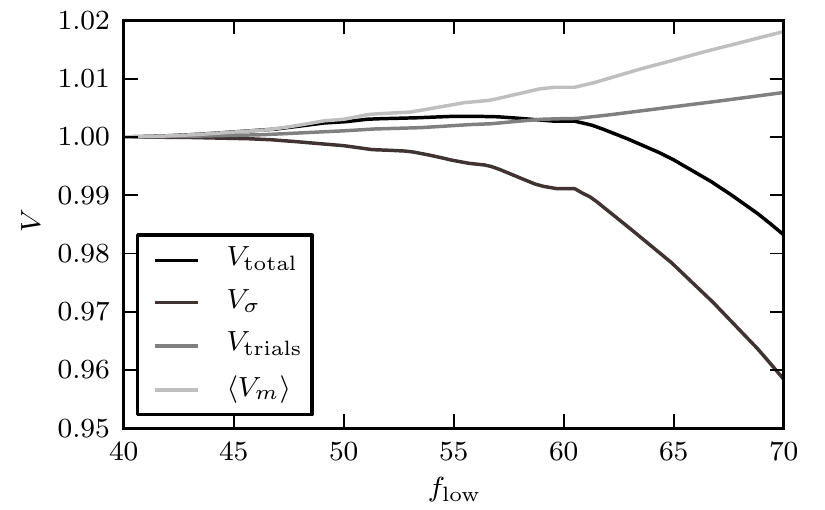}\label{fig:H1S6cost}}
\qquad
\subfloat[V1 VSR2-3 PSD]{\includegraphics[]{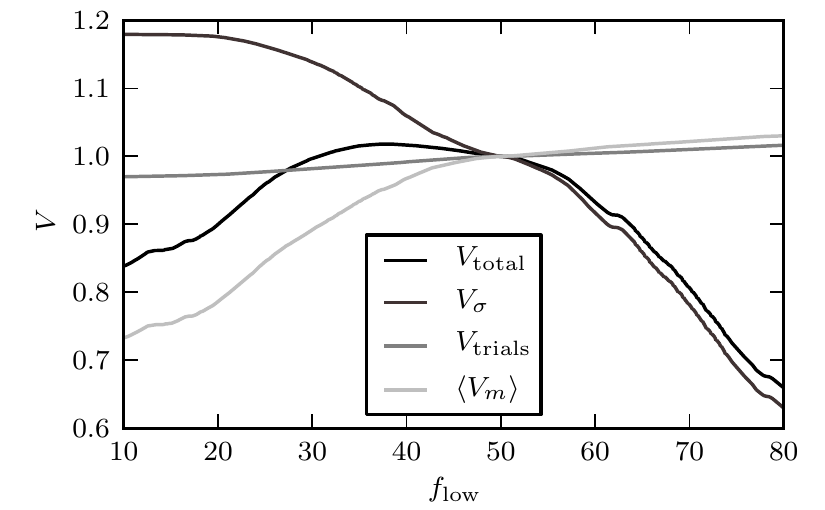}\label{fig:V1VSR23cost}}
\qquad
\subfloat[H1L1V1 S6/VSR2-3 Harmonic Sum PSD]{\includegraphics[]{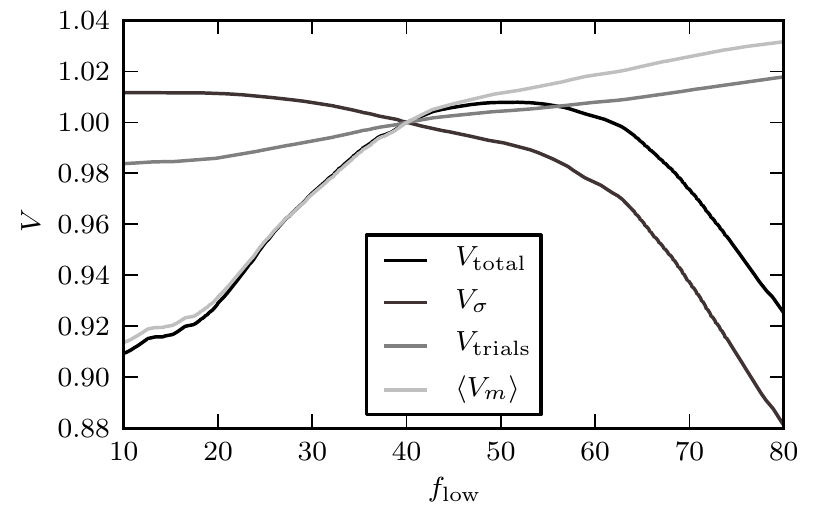}\label{fig:H1L1V1S6VSR23cost}}
\caption{\protect\subref{fig:H1S6cost} shows the constrained average relative volume as
a function of lower frequency cutoff for the H1 LIGO detector during the S6 era.
\protect\subref{fig:V1VSR23cost} and \protect\subref{fig:H1L1V1S6VSR23cost} show the same for
the Virgo detector and H1L1V1 detector network for the VSR2-3 and S6/VSR2-3
eras, respectively. The computational cost of the searches associated with
these eras is given by the fixed cost algorithm. Each panel also contains
traces for the contributions to the constrained average relative volume from
the recoverable \ac{SNR}~$V_{\sigma}$, the trials factor~$V_{\rm trials}$, and
the average template bank mismatch~$\langle V_{m} \rangle$.}
\label{fig:S6VSR23cost}
\end{figure}

\begin{figure}[h]
\centering
\subfloat[H1 \ac{aLIGO} PSD]{\includegraphics[]{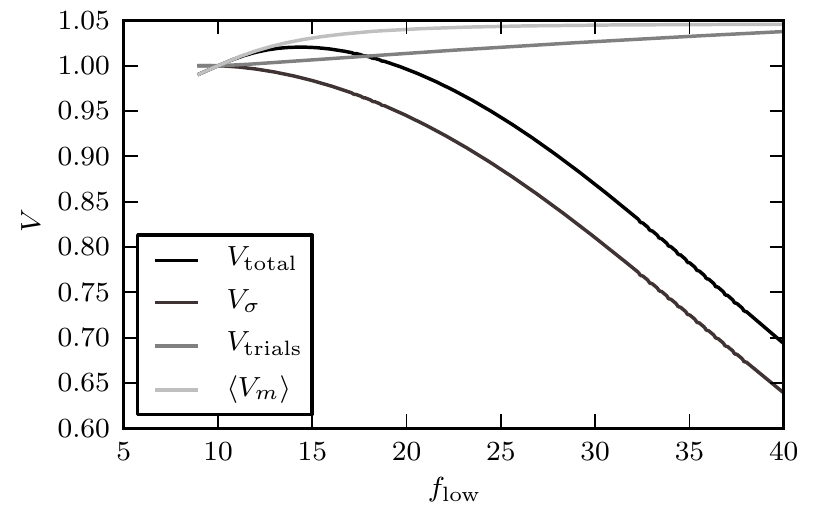}\label{fig:H1aLIGOcost}}
\qquad
\subfloat[V1 \ac{aVirgo} PSD]{\includegraphics[]{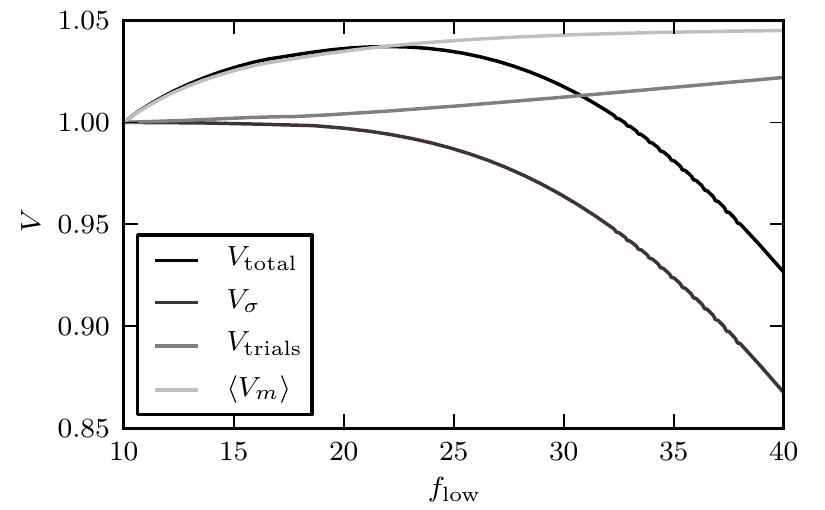}\label{fig:V1aVirgocost}}
\qquad
\subfloat[H1L1V1 \ac{aLIGO}/\ac{aVirgo} Harmonic Sum PSD]{\includegraphics[]{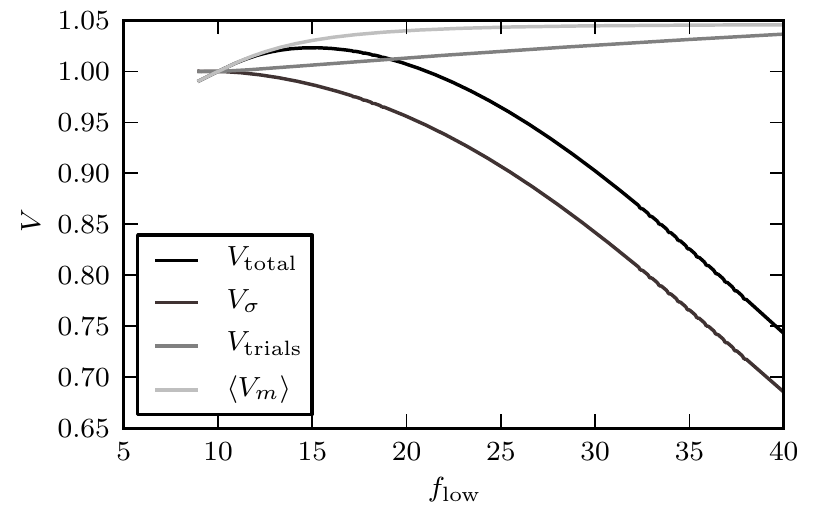}\label{fig:H1L1V2aLIGOaVirgocost}}
\caption{\protect\subref{fig:H1aLIGOcost} shows the constrained average relative volume
as a function of lower frequency cutoff for the proposed H1 LIGO detector
during the \ac{aLIGO} era.  \protect\subref{fig:V1aVirgocost} and
\protect\subref{fig:H1L1V2aLIGOaVirgocost} show the same for the proposed Virgo
detector and H1L1V1 detector network for the \ac{aVirgo} and
\ac{aLIGO}/\ac{aVirgo} eras, respectively. The computational cost of the
searches associated with these eras is given by the cycles cost algorithm,
\eqref{eq:cyclescost}. Each panel also contains traces for the contributions to
the constrained relative average volume from the recoverable
\ac{SNR}~$V_{\sigma}$, the trials factor~$V_{\rm trials}$, and the average
template bank mismatch~$\langle V_{m} \rangle$.}
\label{fig:aLIGOaVirgocost}
\end{figure}

\begin{figure}[h]
\centering
\includegraphics[]{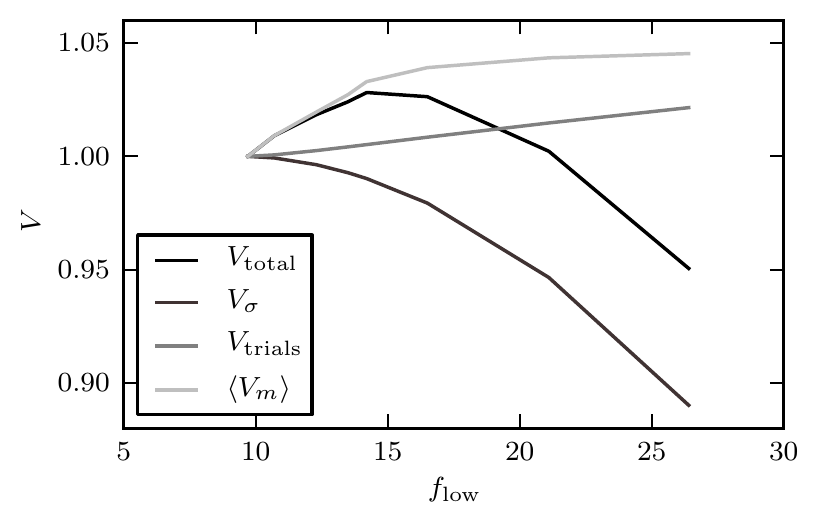}
\caption{We show the constrained average relative volume as a function of low
frequency cutoff for the proposed H1L1V1 detector network during the
\ac{aLIGO}/\ac{aVirgo} era.  The computational cost of this search is given by
the \texttt{LLOID} algorithm.  The contributions to the constrained average
relative volume from the recoverable \ac{SNR} $V_{\sigma}$, the trials factor
$V_{\rm trials}$, and the average template bank mismatch $\langle V_{m}
\rangle$ are also shown. It is interesting to see that the optimal choices for
this search are similar to that of a search where the computational cost is
given by the number of cycles in the template waveform.}
\label{fig:ER1cost}
\end{figure}

As before, we also show a more detailed comparison of the constrained
optimization of the lower frequency cutoff and minimal match as a function of
the lower frequency cutoff. This can be found in
Figs.~\ref{fig:S5VSR1cost}-\ref{fig:ER1cost}. In this situation, the largest
increase in sensitivity is a few percent, coming from the proposed \ac{aVirgo}
detector's \ac{PSD}. Figure~\ref{fig:V1aVirgocost} shows that the majority of
the effect here is coming from decreasing the maximum mismatch (i.e.,
increasing the minimal match) of the template bank from 3\% maximum mismatch to
0.56\% maximum mismatch. In this situation, the drive toward larger lower
frequency cutoffs seems to come from the reduction in the computational cost
per template associated with the total number of cycles contained in the
waveform, as opposed to reducing the trials factor effect.

Finally, we also compare the previous choice of lower frequency cutoff and
minimal match suggested in~\cite{Babak2006} (i.e., $m_{\rm max}=5\%$ and lower
frequency cutoff such that fractional \ac{SNR} loss is 1\%) to the optimal
choice at the same computational cost. This comparison can be found in
Table~\ref{tbl:costoldf_low}. This choice is closer to the optimal choice,
although the optimal choice still provides sensitivity gains as large as one
percent for the \ac{aLIGO}/\ac{aVirgo} detector network.

\FloatBarrier
\section{Conclusion}

We have presented an analysis of the two tunable variables that affect searches
for inspiral signals in \ac{GW} data. We find that with the minimal match of
the template bank held fixed, there is an optimal choice for the lower
frequency cutoff below which reducing this parameter reduces the sensitivity of
a search that employs a maximum likelihood ratio estimate of the \ac{SNR}. This
could be seen as the following inverse result. Even though decreasing the lower
frequency cutoff does not gain significant amounts of \ac{SNR}, it still
provides discriminating power in determining the parameters, thus increasing
the trials factor associated with a fixed region of parameter space.

In addition, through careful balancing of the computational cost associated
with the lower frequency cutoff and the minimal match of the template bank, we
show that improved performance can be achieved at fixed computational cost.
This is the first work that has laid out a procedure for determining the
optimal choice of these parameters for searches for \ac{BNS} \ac{GW} signals
from non-spinning objects.  As searches for inspiral \ac{GW} signals from other
systems can involve additional waveform parameters, and thus larger
computational cost, it will be important to apply this method to other
parameter spaces (e.g., the parameter space of waveforms from binary systems
that including effects from the objects' spins) in order to maximize the
sensitivity of those searches.

\acknowledgments

The author would like to thank Kipp Cannon, Thomas Dent, Chad Hanna, and
Frederique Marion for useful discussions and insightful questions associated
with this work. The author would also like to gratefully acknowledge the US
National Science Foundation, the LIGO Scientific Collaboration, and the Virgo
Collaboration for making public sensitivity curves from LIGO's S5 and S6 and
Virgo's VSR1-3 science runs. The author was supported from the Max Planck
Gesellschaft. This article has the LIGO document number LIGO-P1200098.

\appendix

\bibliography{references}
\end{document}